\documentclass[prl,twocolumn,showpacs,preprintnumbers,amsmath,amssymb]{revtex4}

\usepackage{graphicx}\graphicspath{./graph/}%
\usepackage{dcolumn}
\usepackage{bm}
\usepackage{color}
\usepackage{verbatim}
\usepackage{amssymb}
\usepackage{amsfonts}
\usepackage{latexsym}
\usepackage{epstopdf}

\definecolor{red}{rgb}{1,0,0}
\definecolor{green}{rgb}{0,1,0}
\definecolor{blue}{rgb}{0,0,1}

\newcommand{\be}{\begin{equation}}
\newcommand{\ee}{\end{equation}}
\newcommand{\bea}{\begin{eqnarray}}
\newcommand{\eea}{\end{eqnarray}}
\newcommand{\bdm}{\begin{displaymath}}
\newcommand{\edm}{\end{displaymath}}

\begin{document}


\title{Observation of  Wakefields and  Resonances in Coherent Synchrotron Radiation}

%
\author{B. E. Billinghurst}
\email{brant.billinghurst@lightsource.ca}
\author{J. C. Bergstrom}
\author{C. Baribeau}
\author{T. Batten}
\author{L. Dallin}
\author{T. E. May}
\author{J. M. Vogt}
\author{W. A. Wurtz}
\affiliation{Canadian Light Source Inc., University of Saskatchewan, Saskatoon, SK S7N 2V3, Canada}

\author{R. Warnock}
\email{warnock@slac.stanford.edu}
\affiliation{SLAC National Accelerator Laboratory, Stanford University, Menlo Park, CA 94025, USA \\
Department of Mathematics and Statistics, University of New Mexico, Albuquerque, NM 87131, USA}
\author{D. A. Bizzozero}
\email{dbizzoze@math.unm.edu}
\affiliation{Department of Mathematics and Statistics, University of New Mexico, Albuquerque, NM 87131, USA}
\author{S. Kramer}
\email{skramer@bnl.gov}
\affiliation{ Brookhaven National Laboratory, Upton, NY 11973, USA}
%
\begin{abstract}
We report on high resolution measurements of resonances in the spectrum of coherent synchrotron
radiation (CSR) at the Canadian Light Source (CLS). The resonances permeate the spectrum at wavenumber intervals
of $0.074 ~\textrm{cm}^{-1}$, and are highly stable under changes in the machine setup (energy, bucket filling pattern,
CSR in bursting or continuous mode). Analogous resonances were predicted long ago in an idealized theory as
 eigenmodes of a smooth toroidal vacuum chamber driven by a bunched beam moving on a circular orbit.
A corollary of peaks in the spectrum is the presence of pulses in the wakefield of the bunch at well defined spatial intervals. Through experiments and
further calculations we elucidate the resonance and wakefield mechanisms in the CLS vacuum chamber, which has a fluted form much different from a smooth torus. The wakefield is observed
directly in the 30-110 GHz range by RF diodes, and indirectly by an interferometer in the THz range. The wake pulse sequence found by diodes is less regular than in the toroidal model, and depends on the point of observation, but is accounted for in a simulation of fields in the fluted chamber. Attention is paid to polarization of the observed fields,
and possible coherence of fields produced in adjacent bending magnets. Low frequency wakefield production appears to be mainly local in a single bend, but multi-bend effects cannot be excluded entirely, and could play a role in high frequency resonances. New simulation techniques have been developed, which should be invaluable in further work.
\end{abstract}


\maketitle

\section{Introduction}

A relativistic particle bunch moving along a curved trajectory within a conducting metallic chamber is accompanied by an electromagnetic wake generated by the interaction between the bunch and the chamber.
 When the bunch or a substructure within it is sufficiently short, the particles can act coherently to produce the wake, and the emitted power varies as the square of the current. This implies an enormous intensity compared to the usual incoherent SR at the same frequency. The character of the wake is strongly influenced by the chamber. For example, the  wake of a bunch moving along a circle of radius $R$ between parallel
plates of separation $h$ is weakened due to the shielding action of the induced image charges \cite{murphy}, and as a result the CSR is exponentially suppressed at wavelengths greater than $\lambda_0 \approx 2h(h/R)^{1/2}$, called the CSR shielding cutoff.

The situation changes dramatically when conducting
side walls are added to form a torus \cite{wm}.  Now the spectral distribution of CSR is predicted to have a series
of narrow peaks associated with excitation of resonant modes of the torus, and the shielding cutoff can be at a longer wavelength, depending on the radial beam-to-wall distance. A transfer of power into these modes occurs when the phase velocity of the modes matches the particle velocity.
There is a close analogy to whispering gallery modes in acoustics, which are described by similar mathematics~\cite{pohang}.

The interaction between the field and the bunch can be characterized by an impedance, which we define in a more general sense than usual.
  Suppose that the purely longitudinal current density can be factored as $I(s-ct)f(x,y)$  where $(s,x,y)$ are the usual accelerator coordinates (arc length on a reference trajectory and transverse displacements). A general representation of the $i-$th component of the full field is $E_i(s,x,y,t)=\hspace{-2mm}-\int \exp(2\pi i k(s-ct))Z_i(s,x,y,k)\hat I(k)dk$, where $Z_i$ is the $i$-th component of the impedance.  We define the wavenumber $k=1/\lambda$ without a factor of $2\pi$.
 The position-dependent wakefield, the object of the present study, is the full field regarded as a function of the distance from the bunch, namely
 $\mathcal{E}_i(z,s,x,y)=E_i(s,x,y,t)$ with $z=s-ct$. We study the time dependence of \break $\mathcal{E}_x(s-ct,\mathbf{r})$, and confirm that it depends on the position $\mathbf{r} = \left(s, x, y\right)$.

Peaks in the impedance as a function of $k$ correspond to resonances. It is expected that resonances in $Z_x$ will correlate with those in $Z_s$, as happens in the analytic theory \cite{wm}.
Structures complementary to resonances appear in the wakefield, through the Fourier transform.
 Consider, for example, an idealized torus impedance consisting of a train of resonances uniformly spaced by $\Delta f$ in frequency. The corresponding time domain wakefield consists of a train of localized pulses, equally spaced by the distance $\Delta z$ given by the approximation
\be
\Delta z=\frac{c}{\Delta f}\approx \frac{2}{3}\bigg[\frac{8D^3}{R}\bigg]^{1/2}\ ,              \label{dz}
\ee
where $D$ is the distance from the orbit to the outer wall of the torus. This periodicity is consistent with detailed wakefield calculations by Agoh \cite{agoh}, Stupakov and Kotelnikov \cite{stupakov}, and Zhou {\it et al.} \cite{zhou}, for radiation  in rectangular chambers of various cross sections, usually in a single bend. Particularly noteworthy is the highly localized nature of the wakefield pulses predicted by these calculations.

In this report we present our time domain measurements of the wakefields generated within the vacuum chamber in the bending magnets ($R=7.143$~m, bend angle $\theta=15^\circ$, $h=32$~mm) in the Canadian Light Source (CLS) storage ring. The wakefield following the bunch is indeed found to consist of a sequence of localized pulses, but the pattern is less regular than in the toroidal model, and depends on the point of observation in the vacuum chamber. At the higher frequencies (6~- 15~cm$^{-1}$, 180~- 450~GHz) the wakefield was characterized indirectly using a high resolution Michelson interferometer. At lower frequencies (1.1~- 3.7~cm$^{-1}$, 33~- 110~GHz) backward propagating fields were observed using a series of RF diodes mounted on a forward-viewing port of the vacuum chamber, referred to as backward detectors.
 Note that by their nature our detectors respond only to transverse components of the fields.

\section{Interferometer Measurements}

The Michelson interferometer measures the interference between a direct and a delayed far infrared (FIR) beam, and can observe peaks of intensity when the path length difference is a multiple of the spacing between pulses in the wakefield. The interference pattern is given by $2A_1A_2\left<E_x(t+\tau)E_x(t)\right>$, where $A_1$ and $A_2$ are amplitudes including propagation losses of the two beams in the instrument, $A_iE_x$ is the horizontal electric field, and $\tau$ is the time delay, with $\left<\cdot\right>$ representing the average over $t$ \cite{goodman}. Expressing the field in terms of the horizontal  impedance and averaging over a long time we see that the intensity is proportional to $\int\exp(2\pi ikc\tau)|Z_x(\mathbf{r},k)\hat I(k)|^2 dk$, an even function of $\tau$. The frequency domain current $\hat I(k)$ may contain contributions from several bunches. By contrast the wakefield of interest is proportional to $\int\exp(2\pi ikc\tau)Z_x(\mathbf{r},k)\hat I(k)dk$, but the two quantities should both show peaks in $\tau$ at roughly the same points.

Radiation for the interferometer measurements was produced by operating the  ring at 2.9 GeV, with the momentum compaction adjusted to give a bunch length of a few millimeters. The  CSR was produced in the bursting mode by running at a current of 12~mA distributed uniformly over 3 consecutive bunches. Coherence of the radiation was confirmed by noting a quadratic dependence on the bunch current.

\begin{figure}[htb]
   \centering
   \includegraphics*[width=\linewidth]{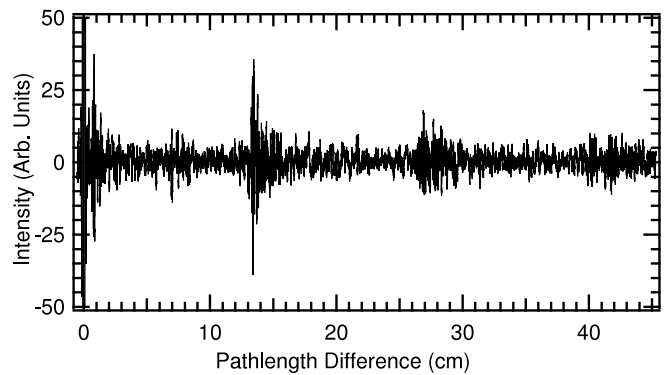}
   \caption{Interferogram as a function of path length difference}
   \label{fig:infero}
\end{figure}

Spectra were acquired on the FIR beamline using a Bruker 125 HR interferometer with a 75~$\mu$m Mylar beam splitter and an Infrared Laboratories Si bolometer as detector  with a gain at 200X.
We used a 12.5~mm aperture, a scanner velocity of approximately 0.051~m/s (80~kHz as measured using a HeNe laser reference), and a resolution of 0.020~cm$^{-1}$ (defined as 0.9 $/$ path-length-difference by Bruker convention). The spectra fall mostly in the frequency range 6-15~cm$^{-1}$ or 0.18-0.45~THz.
The intensity versus path length difference is shown in Fig~\ref{fig:infero}.
The signal comes through the FIR beamline from the M1 mirror at the end of the bend.
Of special note is the strong interference pattern at 13.5~cm and the weaker patterns near 27~cm and 41~cm, which we associate with trailing wakes as discussed in the Introduction.
We have verified that the interference patterns disappear with incoherent synchrotron radiation, which helps to rule out instrumentation anomalies as a factor in the measurements.

The polarization of the signal transmitted by the FIR beamline was analyzed using a wire grid polarizer with a wire spacing of 10~$\mu$m, placed at the entrance to the interferometer. The vertical polarization (as it would appear at the M1 mirror) was found to be negligible, less than 0.5\% of horizontal.

\begin{figure}[htb]
   \centering
   \includegraphics*[width=\linewidth]{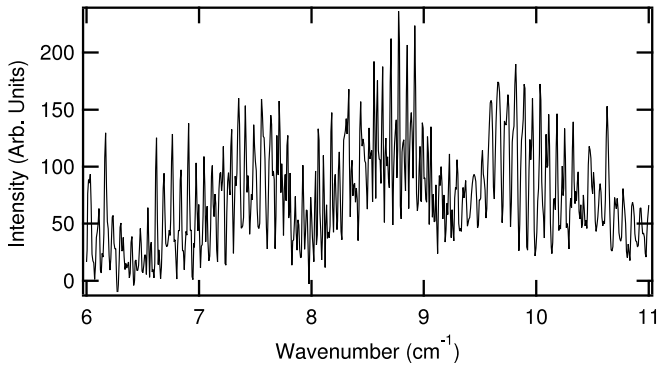}
   \caption{Fourier transform of the interferogram}
   \label{fig:spec}
\end{figure}

Turning now to the Fourier transform of the interferogram, we find the result of Fig.~\ref{fig:spec}. The fine structure is spaced by $\Delta k\approx$ 0.074~cm$^{-1}$, the reciprocal of the 13.5~cm spacing of Fig.~\ref{fig:infero}.
We have accumulated similar data over a period of years, and of particular note is the stability of the spectral pattern of Fig.~\ref{fig:spec}
, including phase, under wide variations in the electron beam parameters such as energy (1.5 or 2.9~GeV), fill pattern (single bunch or 210~bunches), and CSR mode (bursting or continuous). Also, large changes in the  beamline optics were made over the years. A graph showing this invariance appears in Ref.~\cite{pohang}.
 This stability argues in favor of the vacuum chamber as being the dominant factor in shaping the spectrum.

\section{RF Diode Measurements}

\begin{figure}[htb]
   \centering
  \includegraphics*[width=\linewidth]{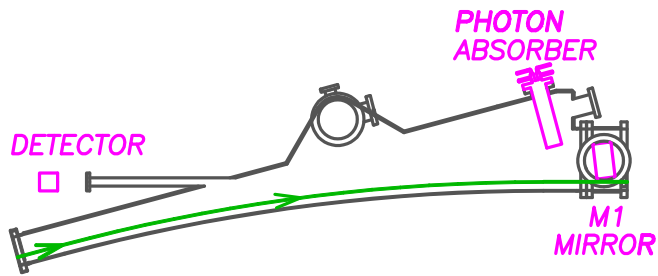}
   \caption{(color online). Fluted vacuum chamber at the FIR dipole with bending radius $R=7.143$~m and deflection angle $\theta=15^\circ$. The maximum excursion of the outer wall from the beam is 33~cm. The diode detector is at the end of the horizontal pipe at the left of the figure.}
   \label{fig:flute}
\end{figure}

The vacuum chambers in the bending magnets are equipped with quartz vacuum windows on 50~cm long tubular extensions mounted near the chamber entrance (Fig.~\ref{fig:flute}).
These align
with the axis of the downstream straight section. RF Schottky diodes (unbiased) mounted outside the windows have proved to be highly sensitive to backward-propagating fields in the 30-110~GHz range. These fields are correlated in time with the bunch and appear to be caused by back-scattering of waves by hardware near the chamber exit. This hardware consists of two parts: a bar-like copper photon absorber which extends about 20~cm into the chamber, at a right angle to the outer wall, and the M1 mirror assembly which directs the FIR radiation upward into the FIR beamline.  Three diodes were used spanning 33-50~GHz, 50-75~GHz (with a 50-75~GHz band-pass filter), and 75-110~GHz \cite{millitech}.
The detector waveguides determine the bandwidths and polarization axes.
To the best of our knowledge, the first application of Schottky diodes to the measurement of freely propagating wakefields is reported in Ref \cite{kramer}.

\begin{figure}[htb]
   \centering
   \includegraphics*[width=\linewidth]{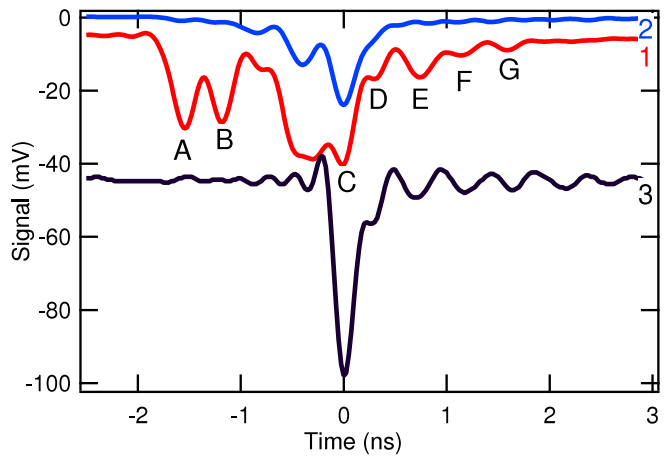}
   \caption{(color online). RF diode measurements in the time domain (oscilloscope traces) with a 50-75~GHz detector. Diode mounting
   and polarization:  1-backward horizontal; 2-backward vertical; 3-forward horizontal (with adjustment of time base).
   For clarity the curves have been separated vertically.}
   \label{fig:diode}
\end{figure}

CSR for the diode work was produced at 2.9~GeV in the continuous mode, with a single bunch of length 2~mm (as deduced from the synchrotron frequency) carrying 1~mA.
 The diodes are highly sensitive to polarization, and we have used this feature to separate the electric fields polarized parallel and perpendicular to the orbital plane, complementing the interferometer study.
The signals from the 50 - 75~GHz diode for these two polarizations are presented in Fig.~\ref{fig:diode} as curves 1 and 2 respectively.
 Our interpretation of the figures is as follows.
The horizontal data (Curve 1) show several downward peaks which we partition into two groups consisting of the first pair (A-B) and the cluster (C-G).
The two groups correspond to reflections from the photon absorber bar and the M1 mirror assembly, respectively, the latter being downstream of the former.
Peaks A and C are interpreted as the prompt signal from the bunch as reflected from the photon absorber and the mirror, respectively. Accordingly, the
A to C distance is 46 cm (1.53 ns), which is twice the physical separation of the two structures (measured center to center).
This interpretation is supported by a test measurement in a bending magnet which had no mirror assembly and only peaks A and B survived.

Peak B is understood as a wake pulse  from the photon absorber following A, whereas D, E, F, G are wake pulses from the mirror following C.
One might ask why there are no further wake pulses from the photon absorber following B, but that is exactly the behavior seen in a simulation reported presently.
Finally, the remaining structures in Curve 1 immediately preceding the peak C correlate with absorbers protecting the mirror assembly and are of no further interest here.

The results for vertical polarization, in Curve 2, show little intensity prior to signals reflected from the mirror assembly. This makes sense, because the mirror assembly is the first structure with strong asymmetry about the midplane $y=0$.  If electromagnetic boundary conditions (and the charge distribution) were symmetrical, the predicted
vertical polarization would be zero.

The RF diode measurements have been repeated at the exit of the FIR beamline, directly in front of the interferometer, and results for horizontal polarization
are shown in Fig.~\ref{fig:diode} as Curve 3. With the time base adjusted to line up the prompt signal (the large peak) with peak C, the subsequent patterns observed at forward and backward detectors line up rather well.  The finer oscillations preceding the big peak in Curve 3 are a manifestation of the approximately 3.5~GHz bandwidth of the oscilloscope.

Although the diode and interferometer measurements are not strictly comparable, operating as they do in different ranges of frequency, the spacing of wakefield pulses as transmitted from the M1 mirror appears to be roughly the same in the two measurements. For instance, the spacing in Curve 3 of  Fig.~\ref{fig:diode}
is about $15\ $cm, versus $13.5\ $cm in Fig.~\ref{fig:infero}.

We turn to the question of wake propagation between bending magnets. Are wakes produced locally or is there cooperation between adjacent dipoles? A cooperative effect was computed in~\cite{stupakov} and~\cite{zhou} for a chamber of constant cross section, but it is parameter dependent and the problem is complicated by our fluted geometry. For an experimental test a different bending magnet was used, one 5.4~m downstream from an adjustable vertical scraper. Otherwise the setup is identical to the FIR dipole, and the backward diode signal is identical to that of Curve~1 of Fig.~\ref{fig:diode}.
The sum of the intensities of leading peaks A and B was measured as a function of vertical aperture over the range 30 - 2~mm.
If a large part of the A + B intensity came from a bend or bends before the scraper, one might expect the intensity to fall markedly when the scraper aperture is at the minimum, at least if the propagating mode were broadly distributed over the open aperture as in the lowest vertical mode in a waveguide.
Instead we found that the A + B intensity \textit{increased} by about 10\%, while the form of the peaks was more or less invariant.
This increase is presumably due to enhanced CSR in the local bend due to a perturbation of the bunch form by the scraper impedance.
This indicates that the A + B wake is produced locally to a considerable extent, since in the straight between the scraper and the local bend there is nothing to directly amplify the radiation.
On the other hand, if only a portion of the final wake was produced upstream and was attenuated by the scraper, the attenuation could be masked by the enhanced local CSR.
Thus we fail to exclude entirely a partial production of the wakefield in upstream magnets.

\section{Simulating the Field Pattern}
 To simulate fields in the fluted chamber we made a time-domain integration of the six Maxwell curl equations, with boundary conditions, by the Discontinuous Galerkin Method \cite{hesthaven}. A Fourier expansion in $y$ leaves two spatial coordinates $(s,x)$. The receding chamber wall is modeled by a flat conducting surface, connected at its end to the photon absorber represented as a rectangle. After the absorber, the wall runs parallel to the beam at $x=33$~cm and then almost perpendicular to the beam to meet the  mirror assembly, which is rendered as a flat surface, tipped by 7.5$^o$ to aim at the center of the bend.
  A Gaussian driving bunch with $\sigma=2~$ mm excites fields in a single pass, the initial condition being the steady state fields in a straight wave guide. A straight exit channel after the mirror is long enough to prevent reflections from the end of the channel reaching the bend region. We used 61195 finite elements of 6th order, 28 nodes per element, for a maximal node spacing of about 1 mm.  The time step was $cdt=0.16~$mm.  The pipe to the diode is not modeled, but the time dependence of the fields  is recorded at the pipe entrance, here denoted as the ``backward port". The diode voltage for horizontal polarization is proportional to $E_x^2$, as conditioned by the detector response. We account for the latter in the frequency domain by the low pass filter $1/(1+if/f_0)$, with $f_0=1~$GHz.

\begin{figure}[htb]
   \centering

   \includegraphics*[width=9cm,height=5cm]{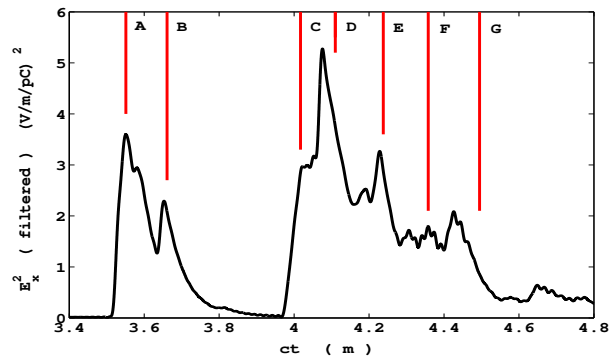}
   \caption{(color online). Simulated $E_x^2$  at backward port {\it vs.} $ct$, after a low pass filter to account for detector response.
   The origin of time $t$ is when the bunch is 5 cm before the entrance to the bend. Only the lowest mode in $y$ is included.}
   \label{fig:sim}
\end{figure}
    The calculated detector response at the backward port shown in Fig.\ref{fig:sim} displays
  signals reflected from the photon absorber and the mirror, just as in the picture inferred from experiment before the simulation.  Field contour plots show sharply defined wave fronts in the reflected fields.   The red lines in Fig.\ref{fig:sim} mark the positions of the experimental peaks in Fig.\ref{fig:diode} (relative to the first theoretical peak). In view of our rudimentary modeling of the reflecting structures, which actually have significant three-dimensional aspects, the resemblance to  the experimental wake pulse positions seems quite satisfactory. The absence of a third wake pulse in the wave from the photon absorber is striking, and emphasizes the dependence of the wake pattern on the point of observation. The calculated pulse D is much too strong and a bit too early, perhaps due to over-simplified modeling of the mirror structure.

  The unfiltered signal shows significant structures between peaks A and B, and between C and D. These might be related to the pulse at 7 cm in the interferogram. 

  The code was validated by checking boundary conditions and by comparison with a well tested paraxial code, which works at points not too close to the end of the bend, before any backward waves are excited.
  A simulation of the interferometer data, lying at higher frequencies, will require higher resolution or possibly a frequency domain method.

 \acknowledgments{
 Research  at the CLS was funded by the Canada Foundation for Innovation, the Natural Sciences and Engineering Research Council of Canada, the National Research Council Canada, the Canadian Institutes of Health Research, the Government of Saskatchewan, Western Economic Diversification Canada, and the University of Saskatchewan. Work at SLAC, University of New Mexico, and Brookhaven was supported by U.S. Department of Energy contracts  DE-AC03-76SF00515, DE-FG02-99ER41104, and DE-AC02-98CH10886
 respectively. We thank Demin Zhou and  Gennady Stupakov for helpful remarks and QMC Instruments for providing the polarizer for these studies.
 }

\end{document}